\documentstyle[aps,floats,epsf,euscript]{revtex}

\begin{document}
\draft
\twocolumn[\hsize\textwidth\columnwidth\hsize\csname@twocolumnfalse%
\endcsname

\flushright{CLNS~04/1867\\
{\tt hep-ph/0403223}\\
March 19, 2004
\vspace{0.2cm}}

\title{Proposal for a Precision Measurement of \boldmath$|V_{ub}|$\unboldmath}

\author{Stefan W. Bosch, Bj\"orn O. Lange, Matthias Neubert, and Gil Paz}

\address{Institute for High-Energy Phenomenology, 
Laboratory for Elementary-Particle Physics\\ 
Cornell University, Ithaca, NY 14853, USA}
\maketitle

\begin{abstract}
A new method for a precision measurement of the CKM matrix element $|V_{ub}|$ is discussed, which combines good theoretical control with high efficiency and a powerful discrimination against charm background. The resulting combined theoretical uncertainty on $|V_{ub}|$ is estimated to be 10\%.
\end{abstract}

]
\narrowtext

\section{Introduction}

One of the main goals of $B$ physics is to perform a detailed study of the flavor sector of the electroweak theory. This can be achieved by accurate measurements of the parameters of the unitarity triangle. The angle $\beta$ has already been determined precisely at the $B$ factories. Because $|V_{ub}|$ is the side opposite to this angle, an accurate measurement of this parameter will directly test the Standard-Model picture of CP violation. It is therefore desirable to device methods for extracting $|V_{ub}|$ with theoretical uncertainties at  the 5--10\% level.

Inclusive semileptonic $\bar B\to X_{u}\,l^{-}\bar\nu$ decays offer the cleanest determination of $|V_{ub}|$. In order to discriminate against the large charm background, one has to apply tight experimental cuts. So far, cuts on the charged-lepton energy $E_{l}>(M_{B}^{2}-M_{D}^{2})/(2M_{B})$, the hadronic invariant-mass squared $s_{H}<M_{D}^{2}$, and the dilepton mass squared $q^{2}>(M_{B}-M_{D})^{2}$ have been employed. Only the $E_{l}$ cut can be applied without neutrino reconstruction. Unfortunately, it has a very low efficiency and is therefore theoretically disfavored. The hadronic-mass cut is in principle the ideal separator between $\bar B\to X_{u} \, l^{-}\bar\nu$  and $\bar B\to X_{c} \,  l^{-}\bar\nu$  events. However, in practice one has to lower the cut due to the experimental resolution on $s_{H}$, thereby reducing the efficiency. In this Letter we elaborate on a highly efficient method for a precision measurement of $|V_{ub}|$ recently proposed in \cite{Bosch:2004th}, which offers several advantages over the existing strategies.

In the relevant region of phase space, differential $\bar B\to X_{u}\,l^{-}\bar\nu$ decay rates can be factorized into perturbatively calculable functions $H_{i}$ and $J_{i}$, and universal shape functions $S_{i}$ \cite{Neubert:1993ch,Bigi:1993ex,Korchemsky:1994jb}. The hard functions $H_{i}$ account for quantum fluctuations at high scales $\mu_{h}\sim m_{b}$, whereas the jet functions $J_{i}$ describe the properties of the final-state hadronic jet at scales $\mu_{i}\sim\sqrt{s_{H}}\sim\sqrt{m_{b}\Lambda_{\rm QCD}}$. Both $H_{i}$ and $J_{i}$ are known to one-loop order in perturbation theory \cite{Bosch:2004th,Bauer:2003pi}. The functions $S_{i}$ parametrize the non-perturbative physics below the intermediate scale $\mu_{i}$. In \cite{Bosch:2004th} we have derived several model-independent constraints on the form of the leading-order shape function $S$. This significantly reduces the hadronic uncertainties in predictions for decay spectra and event fractions. In principle, the shape function can be constrained further using information about the $\bar B\to X_{s}\gamma$ photon spectrum.

\begin{figure}[t]
\epsfxsize=7.35cm
\centerline{\epsffile{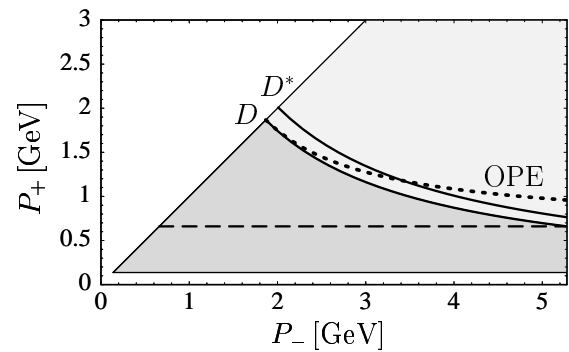}}
\vspace{0.2cm}
\centerline{\parbox{14cm}{\caption{\label{fig:phase}
Phase space for the variables $P_-$ and $P_+$. The different shadings separate the regions where $s_H=P_{+}P_{-}<M_D^2$ (dark gray) and $s_H>M_D^2$ (light gray). The two solid lines correspond to the exclusive decays $\bar B\to D^{(*)}l^{-}\bar\nu$, whereas the dotted line displays the lower boundary for inclusive $\bar B\to X_{c} \, l^{-}\bar\nu$ decays as predicted by the heavy-quark expansion. The dashed line corresponds to $P_+=M_D^2/M_B$.}}}
\end{figure}

The hadronic physics in $\bar B\to X_{u} \, l^{-}\bar\nu$ is most naturally described using the variables $P_{\pm}=E_{H}\mp |\vec P_{H} |$, where $E_{H}$ and $\vec P_{H}$ are energy and three-momentum of the final-state hadronic jet. The phase space in these variables is depicted in Figure \ref{fig:phase} (for $P_+<3$\,GeV). The vast majority of events is located in the ``shape-function region'' of small $P_{+}$ and large $P_{-}$, corresponding to $E_{H}\gg \sqrt{s_{H}}$. In this region a rigorous   theoretical description can be obtained using soft-collinear effective theory \cite{Bauer:2000ew}, which provides a systematic expansion in the small parameter $\Lambda_{\rm QCD}/m_{b}$. All $\bar B\to X_{c} \, l^{-}\bar\nu$ events are confined to the light-shaded region. Cutting on the variable $P_{+}$ allows one to isolate the shape-function region from the rest of phase space. Choosing $P_{+}\le M_{D}^{2}/M_{B}\approx 0.66\,$GeV eliminates the charm background entirely while maintaining a high efficiency. This cut samples the same hadronic phase space as the lepton-endpoint cut $E_{l}\ge 2.31\,$GeV, but it contains significantly more events. Here we propose to employ such a cut to determine $|V_{ub}|$.

The hadronic-mass cut $s_{H}\le M_{D}^{2}$ also fully contains the shape-function region. However, in addition it includes a triangular region of larger $P_{+}$. Since the contour $s_{H}=M_{D}^{2}$ borders the charm background, the observed event fraction is very sensitive to ``leakage'' of charmed events due to imperfect experimental resolution. The $P_{+}$ cut, on the other hand, provides a ``buffer zone'' against charm background.

\section{Calculation of the event fraction}

The fraction $F_P$ of all $\bar B\to X_u\,l^-\bar\nu$ events with hadronic light-cone momentum $P_+\le\Delta_P$ is given by
\begin{eqnarray}\label{FP}
   F_P(\Delta_P)
   &=& T(a)\,e^{V_H(\mu_{h},\mu_i)} \left( \frac{m_b}{\mu_h} \right)^{-a}
    \int_0^{\Delta_P}\!d\hat\omega\,\hat S(\hat\omega,\mu_i) \nonumber\\
   &&\hspace{-1.5cm}\times \bigg\{ 1 + \frac{\alpha_s(\mu_{h})}{3\pi} \Big[
    - 4\ln^{2}\frac{m_{b}}{\mu_{h}} + H_{1}(a) \ln\frac{m_{b}}{\mu_{h}}
    + H_{2}(a) \Big] \nonumber\\
   &&\hspace{-0.77cm}\mbox{}+ \frac{\alpha_s(\mu_i)}{3\pi} \Big[ 2L_\omega^2
    + J_1(a)\,L_\omega + J_2(a) \Big] \bigg\} \,, \\[-0.5cm] \nonumber
\end{eqnarray}
where
\begin{equation}
   a = \frac{16}{25}\,\ln\frac{\alpha_s(\mu_i)}{\alpha_s(\mu_{h})} \,, \qquad
   L_\omega = \ln\frac{m_b(\Delta_P-\hat\omega)}{\mu_i^2} \,,
\end{equation}
and the functions $T$, $V_{H}$, $H_{i}$, and $J_i$ can be found in~\cite{Bosch:2004th}. This result is valid in the shape-function region $\Delta_{P}\sim\Lambda_{\rm QCD}$ and at leading power in the heavy-quark expansion. Subleading contributions can be calculated systematically using the techniques developed in \cite{Bosch:2004th}. Large logarithms have been resummed into a leading-order Sudakov factor $T(a) \,e^{V_{H}}$ and next-to-leading order functions $H_{i}$ and $J_{i}$. The result for $F_P$ is formally independent of the two matching scales $\mu_{h}$ and $\mu_{i}$.

The fraction $F_{M}$ of events with hadronic invariant mass such that $s_{H}\le s_{0}$ receives a contribution from the box-shaped region below the dashed line in Figure \ref{fig:phase} as well as the triangular dark-shaded region above that line. The former simply yields $F_{P}(\Delta_{s})$ with $\Delta_{s}=s_{0}/M_{B}$, while the latter involves a weighted integral over the shape function up to $\sqrt{s_{0}}$ \cite{Bosch:2004th}. Unlike for the $P_{+}$ spectrum, for which knowledge of the shape function in the range $0\le\hat\omega\le\Delta_{P}\sim\Lambda_{\rm QCD}$ is needed, here the functional form of the shape function is required over a wider range. Because the collinear expansion breaks down near the tip $P_{+}\sim P_{-}\sim\sqrt{s_{0}}$ of the triangular region, it is an open challenge how to construct a systematic heavy-quark expansion for $F_{M}(s_{0})$.

Our prediction for the fraction of events with the optimal cut $P_{+}\le M_{D}^{2}/M_{B}$ is
\begin{equation}\label{FPnum}
   F_{P} = (79.6\pm 10.8\pm 6.2\pm 8.0)\% \,,
\end{equation}
where the errors represent the sensitivity to the shape function, an estimate of ${\cal O}(\alpha_{s}^{2})$ contributions, and power corrections, respectively. The uncertainty due to our ignorance of the shape function is obtained by varying the parameters $\bar\Lambda=M_{B}-m_{b}$ and $\mu_{\pi}^{2}$ determining its first two moments ($\delta F_{P}=\phantom{}^{+8.2}_{-8.1}\%$), as well as the functional form of $S$ ($\delta F_{P}=\phantom{}^{+6.3}_{-7.8}\%$). Here, $m_{b}$ and $\mu_{\pi}^{2}$ are defined in the shape-function scheme \cite{Bosch:2004th}. Higher-order perturbative effects are estimated by studying the dependence on the matching scales $\mu_{i}$ and $\mu_{h}$, varied in the ranges $1.25\,{\rm GeV}\le\mu_{i}\le 1.75\,{\rm GeV}$ ($\delta F_{P}=\phantom{}^{+4.7}_{-4.9}\%$) and $m_{b}/\sqrt{2}\le\mu_{h}\le\sqrt{2}\,m_{b}$ ($\delta F_{P}=\phantom{}^{+4.1}_{-3.9}\%$). The perturbative uncertainty could be reduced by computing the ${\cal O}(\alpha_{s}^{2})$ corrections to Eq.~(\ref{FP}). Power corrections will be discussed in more detail below. 

The CKM-matrix element $|V_{ub}|$ can be extracted by comparing a measurement of the partial rate $\Gamma_{u}(P_{+}\le\Delta_{P})$ with a theoretical prediction for the product of the event fraction $F_{P}$ and the total inclusive $\bar B\to X_{u} \, l^{-}\bar\nu$ rate. The resulting theoretical uncertainty on $|V_{ub}|$ is
\begin{equation}
   \frac{\delta |V_{ub}|}{|V_{ub}|} = (\pm 7\pm 4\pm 5\pm 4)\% \,,
\end{equation}
where the last error comes from the uncertainty in the total rate \cite{Hoang:1998ng,Uraltsev:1999rr}. Because of the large efficiency of the $P_{+}$ cut, weak annihilation effects \cite{Voloshin:2001xi} have an influence on $|V_{ub}|$ of less than 2\% and can be safely neglected.

At leading power in $\Lambda_{\rm QCD}/m_{b}$ the shape-function uncertainty could be eliminated by relating the $P_{+}$ spectrum to the $\bar B\to X_{s}\gamma$ photon spectrum, both of which are given (at tree level) directly by the shape function. A prototype of such a relation can be found in \cite{Bosch:2004th}, where the charged-lepton energy spectrum was expressed as a weighted integral over the $P_{+}$ spectrum. Using similar methods, it will be possible to construct a shape-function independent relation of the form
\begin{equation}
   F_{P}(\Delta) = \frac{1}{\Gamma_{s}} 
   \int_{\frac{M_{B}-\Delta}{2}}^{\frac{M_{B}}{2}}\!dE_{\gamma}\,
   \frac{d\Gamma_{s}}{dE_{\gamma}}\,w(\Delta,E_{\gamma}) \,,
\end{equation}
where at tree level the weight function is simply $w(\Delta,E_{\gamma})=1$. At next-to-leading order the energy dependence of this function resides in a single logarithm $\alpha_{s}(\mu_{i})\ln[m_{b}(\Delta-M_{B}+2E_{\gamma})/\mu_{i}^{2}]$. A similar expression relating the hadronic-mass spectrum with the $\bar B\to X_{s}\gamma$ photon spectrum would be far more complicated and require input of the photon spectrum beyond the region where it is currently experimentally accessible.

In order to study power corrections it is instructive to resort to a local operator-product expansion (OPE), which is valid for large values of $P_{+}$ and $P_{-}$. The resulting prediction for the normalized $\bar B\to X_u\,l^-\bar\nu$ spectrum in the variable $\hat p_+=p_{+}/m_{b}=(P_{+}-\bar\Lambda)/m_b$ reads
\begin{eqnarray}
   \frac{1}{\Gamma_u}\,\frac{d\Gamma_u}{d\hat p_+}
   &=& \left( 1 - \frac{463}{36}\,\frac{\alpha_s}{3\pi} \right)
    \delta(\hat p_+) \nonumber\\
   &+& \frac{\alpha_s}{3\pi} \left[
    - 4 \left( \frac{\ln\hat p_+}{\hat p_+} \right)_{\!*}
    - \frac{26}{3} \left( \frac{1}{\hat p_+} \right)_{\!*}
    + h(\hat p_+) \right] \nonumber\\
   &-& \left( \frac{17\lambda_1}{18 m_b^2}
    + \frac{3\lambda_2}{2m_b^2} \right) \delta'(\hat p_+)
    - \frac{\lambda_1}{6m_b^2}\,\delta''(\hat p_+) \,,
\end{eqnarray}
where $0\le\hat p_+\le 1$, and
\begin{eqnarray}
   h(p)
   &=& \frac{158}{9} + \frac{407p}{18} - \frac{367p^2}{6}
    + \frac{118p^3}{3} - \frac{100p^4}{9} \nonumber\\
   &+& \frac{11p^5}{6} - \frac{7p^6}{18} 
    - \left( \frac{4}{3} - \frac{46p}{3} - 6p^2
    + \frac{16p^3}{3} \right) \ln p \nonumber\\
   &-& 4p^2(3-2p) \ln^2 p \,.
\end{eqnarray}
We include the contributions from dimension-3 operators at $O(\alpha_s)$ and those from dimension-5 operators (whose matrix elements are proportional to the heavy-quark effective theory parameters $\lambda_{1,2}$) at tree level, using \cite{Manohar:1993qn,DeFazio:1999sv}. If the $\hat p_+$ spectrum is integrated without a weight function, the tree-level power corrections from dimension-5 operators vanish. This is in accordance with the fact that subleading shape functions have zero norm at tree level \cite{Bauer:2002yu}. For $\Lambda_{\rm QCD}\ll\Delta_P\ll m_b$ we obtain with $\overline{\Delta}=\Delta_P-\bar\Lambda$
\begin{eqnarray}\label{FPOPE}
   F_P(\Delta_P) = 1 &-& \frac{\alpha_s}{3\pi} \bigg[
    \left( 2\ln^2\frac{m_b}{\overline{\Delta}}
    - \frac{26}{3}\ln\frac{m_b}{\overline{\Delta}} + \frac{463}{36} \right)
    \nonumber\\
   &-& \frac{\overline{\Delta}}{m_b} \left(
    \frac{4}{3}\,\ln\frac{m_b}{\overline{\Delta}} + \frac{170}{9} \right)
    \bigg] + \dots \,.
\end{eqnarray}
For $\Delta_{P}\sim 1$\,GeV, the power correction in the second line leads to an enhancement of the fraction $F_{P}$ by 5--10\%. This is of similar magnitude as tree-level estimates of (zero-norm) subleading shape-function effects on the $E_{l}$ and $s_{H}$ spectra \cite{Neubert:2002yx,Burrell:2003cf}.

The leading-order term in Eq.~(\ref{FPOPE}) can also be obtained from Eq.~(\ref{FP}) by taking the limit $\Delta_{P}\gg\Lambda_{\rm QCD}$. Interestingly, such an analysis uncovers that in this kinematic range there is an enhanced class of power corrections of the form $(\Lambda_{\rm QCD}/\overline{\Delta})^n$ with $n\ge 2$, which arise first at $O(\alpha_s)$ (see also \cite{Bauer:2003pi}). The leading corrections to the expression above are given by
\begin{equation}
   \delta F_P(\Delta_P) = - \frac{\alpha_s}{3\pi}
   \left( 2\ln\frac{m_b}{\overline{\Delta}} - \frac{7}{3} \right)
   \frac{\mu_\pi^2}{3\overline{\Delta}^2} \,.
\end{equation}
It follows that for sufficiently large $\overline{\Delta}$ the event fraction can be expressed as a double expansion in $\overline\Delta/m_{b}$ and $\Lambda_{\rm QCD}/\overline{\Delta}$. Such a multi-scale OPE is non-trivial and left for future work.

\section{Charm background}

One of the main advantages of the $P_{+}$ spectrum over the hadronic-mass spectrum is a better control of the charm background. In order to study this in more detail, we investigate the OPE prediction for the normalized $\hat p_+$ spectrum in $\bar B\to X_c\,l^-\bar\nu$ decays,
\begin{eqnarray}\label{charmOPE}
   \frac{1}{\Gamma_c}\,\frac{d\Gamma_c}{d\hat p_+}
   &=& \frac{2(\varrho-\hat p_+^2)^2}{f(\varrho)\,\hat p_+^5}
    \Big[ \hat p_+^3 (3 - 2\hat p_+) \nonumber\\
   &&\!\mbox{}+ \varrho \,\hat p_+ (3 - 8\hat p_+ + 3\hat p_+^2)
    - \varrho^2 (2 - 3\hat p_+) \Big] \,,
\end{eqnarray}
where $\varrho\le\hat p_+\le\sqrt{\varrho}$ with $\varrho=(m_c/m_b)^2$, and
\begin{equation}
   f(\varrho) = 1 - 8\varrho + 8\varrho^3 - \varrho^4
   - 12\varrho^2\ln\varrho \,.
\end{equation}
We only include tree-level contributions from dimension-3 operators. In this approximation, inclusive charm events are located along the dotted line in Figure~\ref{fig:phase}. Because ${\cal O}(\alpha_{s})$ corrections redistribute these events into the light-gray segment above that line, the integral over the spectrum in Eq.~(\ref{charmOPE}) serves as an upper bound on the inclusive charm background. In the same approximation, the hadronic mass distribution is given by
\begin{eqnarray}\label{FMOPE}
   \frac{1}{\Gamma_c}\,\frac{d\Gamma_c}{d\hat s}
   &=& \frac{2}{f(\varrho)\,\varepsilon} \sqrt{z^2-4\varrho}\,
    \Big[ z(3-2z) - \varrho(4-3z) \Big] \,, \nonumber\\
   \mbox{with~~}
   z &=& \frac{\hat s-\varepsilon^2-\varrho}{\varepsilon} \,.
\end{eqnarray}
Here, $\hat s=s_H/m_b^2$, $\varepsilon=\bar\Lambda/m_b$, and phase space is such that $(\sqrt{\varrho}+\varepsilon)^2\le\hat s \le(\varrho+\varepsilon)(1+\varepsilon)$.

\begin{figure*}[t]
\begin{center}
\epsfxsize=15.0cm
\epsffile{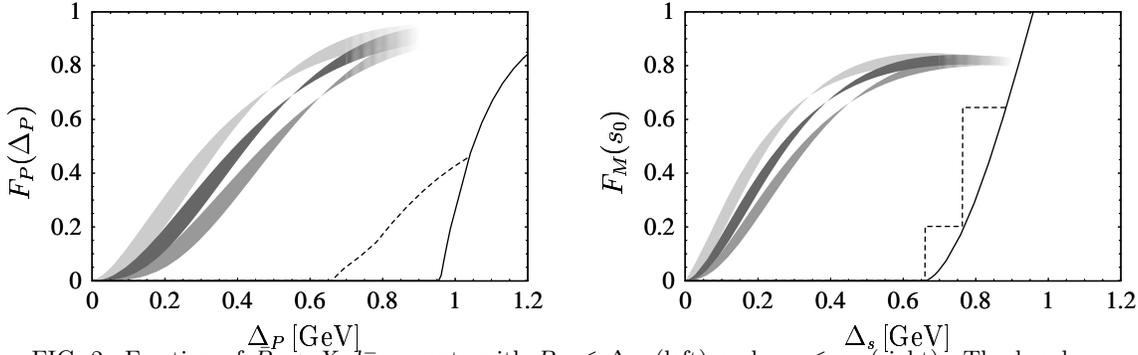}
\vspace{-0.3cm}
\parbox{14cm}{\caption{\label{fig:Fcharm}
Fraction of $\bar B\to X_{u}\,l^{-}\nu$ events with $P_{+}\le\Delta_{P}$ (left) and $s_{H}\le s_{0}$ (right). The bands correspond to the values $\bar\Lambda=0.63$\,GeV (dark), 0.70\,GeV (medium), and 0.56\,GeV (light). Their width reflects the sensitivity to the value of $\mu_\pi^2$ varied between 0.20 and 0.34\,GeV$^2$. The curves represent the background from inclusive $\bar B\to X_{c}\,l^-\bar\nu$ (solid) and exclusive $\bar B\to D^{(*)} l^-\bar\nu$ decays (dashed), normalized to the total inclusive semileptonic charm rate.}}
\end{center}
\end{figure*}

The results for the event fractions $F_{P}$ and $F_{M}$ are summarized in Figure~\ref{fig:Fcharm}. The bands show the inclusive $\bar B\to X_u\,l^-\bar\nu$ event fractions for different shape-function models \cite{Bosch:2004th}, while the solid lines give predictions for the inclusive $\bar B \to X_c\, l^- \bar \nu$ background spectra as obtained from Eqs.~(\ref{charmOPE}) and (\ref{FMOPE}). The latter are normalized to their total decay rate $\Gamma_c$, which is about 60 times larger than $\Gamma_u$. In the case of the fraction $F_{M}$, the inclusive charm background starts right at the threshold $\Delta_{s}=(m_{c}+\bar\Lambda )^{2}/M_{B}\simeq M_{D}^{2}/M_{B}$. In the case of the $F_{P}$ event fraction, on the contrary, the inclusive charm background starts at a value $\Delta_{P}=m_{c}^{2}/m_{b}+\bar\Lambda\approx 0.96\,{\rm GeV}$, which is significantly larger than the value $\Delta_{P}=M_{D}^{2}/M_{B}\approx 0.66\,{\rm GeV}$ above which final states containing charm mesons are kinematically allowed. This is a reflection of the fact that there is a gap between the dashed line representing the $P_{+}$ cut and the dotted curve representing the onset of the continuum of charmed final states in Figure~\ref{fig:phase}. In reality this gap is filled by exclusive modes containing the hadronic final states $D$, $D^*$, or $D \pi$, $D\pi\pi$. The exclusive contributions to the hadronic $P_+$ spectrum from $\bar B\to D^{(*)} l^-\bar\nu$ decays are
\begin{eqnarray}
   \frac{d\Gamma_D}{dP_+}
   &=& \frac{G_F^2|V_{cb}|^2 M_B^5}{48\pi^3}\,
    \frac{(1+r)^2\,r^3}{P_+}\,(w^2-1)^2\,|{\cal F_D}(w)|^2 \,, \nonumber\\
   \frac{d\Gamma_{D^*}}{dP_+}
   &=& \frac{G_F^2|V_{cb}|^2 M_B^5}{48\pi^3}\,
    \frac{(1-r_*)^2\,r_*^3}{P_+}\,(w^2-1)\,(w+1)^2 \nonumber\\
   &\times& \left( 1 + \frac{4w}{w+1}\,
    \frac{1 - 2w r_* + r_*^2}{(1-r_*)^2} \right)
    |{\cal F_{D^*}}(w)|^2 \,,
\end{eqnarray}
where $M_{D^{(*)}}^2/M_B\le P_+\le M_{D^{(*)}}$ and $r_{(*)}=M_{D^{(*)}}/M_B$. The recoil variable $w = v\cdot v'$ is given by $M_{D^{(*)}}^{2}+P_{+}^{2}=2w\,M_{D^{(*)}}P_{+}$. The corresponding contributions to the hadronic mass spectrum are given by
\begin{equation}
   \frac{d\Gamma_{D^{(*)}}}{ds_H} = \Gamma_{D^{(*)}} \cdot
   \delta(s_H-M_{D^{(*)}}) \,.
\end{equation}
The exclusive contributions to the event fractions are given by the dashed lines in Figure~\ref{fig:Fcharm}. To obtain these curves we have used an ansatz for the form factors ${\cal F}_{D^{(*)}}(w)$ that is consistent with experimental data on the recoil spectra and branching fractions. The fact that these exclusive modes are very well understood should help to model the background. The smooth onset of the $D^{(*)}$ background is a direct consequence of the fact that the ideal $P_+$ cut touches the charm region at only a single point in phase space (see Figure \ref{fig:phase}). On the contrary, the region of phase space when applying the $s_{H}$ cut borders the background along the curve separating the light- and dark-shaded regions, which leads to a step increment in the event fraction $F_{M}$. As a consequence, one needs to move away from the ideal cut $s_H=M_D^2$ because of smearing effects due to experimental resolution. It is our hope that it will be possible to stay closer to the ideal cut when performing a $P_+$ analysis. In this case, both the $P_+$ and $s_H$ discrimination methods lead to comparable efficiencies and shape-function uncertainties.

\section{Conclusions}

We have presented a new method for a precision measurement of the CKM matrix element $|V_{ub}|$ in inclusive $\bar B\to X_{u}l^{-}\nu$ decays, using a cut on the hadronic light-cone variable $P_{+}=E_{H}-|\vec P_{H}|$. This method has been compared to a strategy using a cut on hadronic invariant mass. While both approaches are highly efficient and comparable in their sensitivity to non-perturbative physics, a measurement of the $P_+$ spectrum is favored in view of its theoretical potential. The reasons are twofold: First, it offers a simpler construction of shape-function independent relations to the photon spectrum in $\bar B \to X_s \gamma$, thus allowing for a theoretically clean determination of $|V_{ub}|$. Secondly, the $P_+$ spectrum can be evaluated within a systematic framework, which makes the calculation of power corrections to Eq.~(\ref{FP}) feasible. This will be necessary in order to lower the theoretical uncertainty on $|V_{ub}|$ to the 5\% level. Such an ambitious goal might also require to include at least part of the ${\cal O}(\alpha_{s}^{2})$ corrections.

\medskip
{\it Acknowledgment:} This research was supported by the National Science Foundation under Grant PHY-0098631.

\end{document}